\documentstyle[12pt,amstex,amssymb,righttag]{article}

\def\llsymbol#1{\@llsymbol{\@nameuse{c@#1}}}
\def\@llsymbol#1{\ifcase#1\or {}\or {'}\or {''}\or {'''}\or 
   {''''}\or {'''''}\or  \else\@ctrerr\fi\relax}

\newcounter{contador}
\newcommand{\letra}{
   \stepcounter{equation}
   \setcounter{contador}{\value{equation}}
   \setcounter{equation}{0}
   \renewcommand{\theequation}{\thecontador.\alph{equation}}}
\newcommand{\antiletra}{
   \renewcommand{\theequation}{\arabic{equation}}
   \setcounter{equation}{\value{contador}}}

\def\citet{\@ifnextchar [{\@tempswatrue\@citey}{\@tempswafalse\@citey[]}}
 
\def\@citey[#1]#2{\if@filesw\immediate\write\@auxout{\string\citation{#2}}\fi
  \def\@citec{}\@cite{\@for\@cited:=#2\do
    {\@citec\def\@citec{--}\@ifundefined
       {b@\@cited}{{\bf ?}\@warning
       {Citation `\@cited' on page \thepage \space undefined}}%
\hbox{\csname b@\@cited\endcsname}}}{#1}}
\begin{document}

\thispagestyle{empty}

\vspace{3cm}

\begin{center}
{\Large \bf Black Hole entropy in $D=2+1$ dimensions from extended
Chern-Simons term in a gravitational background}\\[0.5cm]

\vspace{3mm}
by

\vspace{3mm}
{\sl Carlos Pinheiro$^{\ast}$}\footnote{fcpnunes@@cce.ufes.br/maria@@gbl.com.br} \\[0.5cm]
and \\[0.5cm]
{\sl F.C. Khanna$^{\ast \ast}$}\footnote{khanna@@phys.ualberta.ca}

\vspace{3mm}
$^{\ast}$Universidade Federal do Esp\'{\i}rito Santo, UFES.\\
Centro de Ci\^encias Exatas\\
Av. Fernando Ferrari s/n$^{\underline{0}}$\\
Campus da Goiabeiras 29060-900 Vit\'oria ES -- Brazil.\\

$^{\ast \ast}$Theoretical Physics Institute, Dept. of Physics\\
University of Alberta,\\
Edmonton, AB T6G2J1, Canada\\
and\\
TRIUMF, 4004, Wesbrook Mall,\\
V6T2A3, Vancouver, BC, Canada.
\end{center}

\vspace{3mm}
\begin{center}
Abstract
\end{center}

We study the contribution to entropy of Black Holes in $D=2+1$
dimensions 
from an extension of the Chern Simons theory including higher derivative
in a curved space-time \cite{dois}.

\newpage
\setcounter{page}{1}
\newpage
\subsection*{Introduction}

\paragraph*{}
The topologicaly theory like Chern-Simons in $D=2+1$ dimensions has
been studied in various different approaches in quantum Field theory, in
particularly in perturbative quantum gravity \cite{seis}.

In general, topological action such as 
$\varepsilon^{\mu \nu \alpha}A_{\mu}\partial_{\nu}A_{\alpha}$ where $A_{\mu}$ means
the abelian potential vector, or the action 
$\varepsilon^{\mu \nu \alpha}\left(\Gamma_{\mu \beta}^{\lambda}\partial_{\nu}
\Gamma_{\alpha \lambda}^{\beta}+\displaystyle{\frac{2}{3}}\ \Gamma_{\mu \lambda}^{\sigma}
\Gamma^{\lambda}_{\nu \gamma}\Gamma^{\gamma}_{\alpha \sigma}\right)$
with $\Gamma^{\mu}_{\alpha \beta}$ being the connection, that do not
contribute to the entropy of  black holes in $D=2+1$. Furthermore
these terms do not contribute dynamically  in a quantum field theory
\cite{seis}. 
Rich physics can be explored in quantum field theory
\cite{seis,dois} when Chern-Simons terms is combined with Maxwell or
Einstein Hilbert Lagrangian.
The extension of Chern-Simons theory including highest derivative in
flat space time or in curved space time is carried out by Jackiw
and Deser \cite{dois}.
The higher derivative Chern-Simons extensions has a strong dependence
on the local field strengh, $F_{\mu \nu}$, and not on the vector
potential, thus the gauge information can be lost \cite{dois}.

On the other hand ``extensions'' such as the usual Chern-Simons
term do not contribute to any change in the original value of
entropy for black holes in $D=2+1$ dimensions \cite{tres,quatro}. In
contrast to that, with the extension of Chern-Simons term in a
gravitational background \cite{dois} some interesting things
happen. We intend to find contributions to entropy of black holes in
$D=3$ dimensions.

Introducing the $I_{ECS}$ (extension for Chern-Simons with Higher
Derivative in a gravitational background) and applying the same
procedure as in \cite{tres} we compute contributions to entropy
of black holes in $D=3$ and we find the inverse e
Hawking evaporation temperature, partition function and stress
energy-momentum tensor. Although the $I_{ECS}$ is not globally
topological \cite{dois} due to its energy-momentum tensor 
$T^{\mu \nu}_{ECS}$ being different from zero the contribution from
$I_{ECS}$ to entropy of black holes can be computed.

The non Abelian case will be treated in the next letter again without
any linkage with topology associated, with metric in accordance
with \cite{cinco}.

Let us begin by writing the funcional integral
\begin{equation}
Z=\int {\cal D}g\ e^{-(I+I_{ECS})}
\end{equation}
where $I$ and $I_{ECS}$ are respectively the action for the three
dimensional gravity with a negative cosmological constant 
$\Lambda =-\displaystyle{\frac{2}{\ell^2}}$ and action for higher derivative
Chern-Simons extension in a curved space time given by
\begin{eqnarray}
I &=& \frac{1}{16\pi G}\int \left(R +\frac{2}{\ell^2}\right)dx^3\quad
\mbox{and}\\
&& \nonumber \\
I_{ECS} &=& -(2m)^{-1}\int \varepsilon^{\alpha \beta \gamma}f_{\alpha}
\partial_{\beta}f_{\gamma}dx^3
\end{eqnarray}
with $f_{\alpha}$ written as
\begin{equation}
f_{\alpha} =\frac{1}{\sqrt{g}}\ g_{\alpha \beta}
\varepsilon^{\beta \mu \nu}\partial_{\mu}A_{\nu}\ . 
\end{equation}
In according with \cite{dois} $f_{\alpha}$ is a covariant vector and
$f^{\alpha}$ being contravariant vector. The metric dependence in
$I_{ECS}$ is completely contained in $f_{\alpha}$.

The equations of motion derived from this action (2) are solved
\cite{um,tres} for the three-dimensional black hole whose metric is
\begin{equation}
ds^2=-\left(-8MG+\frac{r^2}{\ell^2}\right)dt^2+
\left(-8MG+\frac{r^2}{\ell^2}\right)^{-1}dr^2+r^2d\varphi^2
\end{equation}
where the quantities $R,\ \varepsilon^{\alpha \beta \gamma},\ 
A_{\mu}(x)$ are the scalar curvature, the Levi-Civita tensor 
$\varepsilon^{012}=+1$ and the vector potential respectively.

The three components of $f_{\alpha}$ are  
\begin{eqnarray}
f_0 &=& \frac{g_{00}}{\sqrt{-g}}\ \varepsilon^{012}
\left(\partial_1A_2-\partial_2A_1\right)\nonumber \\
f_1 &=& \frac{g_{11}}{\sqrt{-g}}\ \varepsilon^{102}
\left(\partial_0A_2-\partial_2A_0\right) \\
f_2 &=& \frac{g_{22}}{\sqrt{-g}}\ \varepsilon^{201}
\left(\partial_0A_2-\partial_1A_0\right)\nonumber 
\end{eqnarray}

On considering the antisymmetry of the Levi-Civita tensor,
the action $I_{ECS}$, can be written as 
\begin{equation}
I_{ECS} \sim \int d^3x\left[f_0\left(\partial_1f_2-\partial_2f_1\right)-
f_1\left(\partial_0f_2-\partial_2f_0\right)+f_2
\left(\partial_0f_1-\partial_1f_0\right)\right]\ .
\end{equation}
We recall that in $D=3$ we have
\[
A_{\mu}=A_{\mu}(x^{\alpha})=(A_0,A_i)=(\varphi ,A_i) \quad i=1,2
\]
and 
\[
x^{\alpha}=(x^0,x^1,x^2)=(t,r,\varphi )
\]
and that the electric and magnetic field are pseudo vector and
scalar respectively.

Thus, we introduce definitions for magnetic and electric fields as 
\begin{eqnarray}
\vec{E} &=& -\vec{\nabla}\varphi +\frac{\partial \vec{A}}{\partial t} \nonumber \\
\vec{B} &=& \vec{\nabla}\times \vec{A}\ .
\end{eqnarray}
Then $f_{\alpha}$ are given as
\begin{eqnarray}
f_0 &=& \frac{g_{00}}{\sqrt{-g}}\ B\ , \nonumber \\
f_1 &=& -\ \frac{g_{11}}{\sqrt{-g}}\ E_{\varphi}\quad \mbox{and} \\
f_2 &=& \frac{g_{22}}{\sqrt{-g}}\ E_r \nonumber 
\end{eqnarray}
where $E_{\varphi}$ and $E_r$ are components of the electric field
and $B$ is the magnetic field in a gravitational background.

Then  the ``Chern-Simons action'' as a function of the metric, electric
and magnetic fields is 
\letra
\begin{eqnarray}
&{}& \frac{g_{00}}{\sqrt{-g}}\ B\left[E_r+r\left(
\frac{\partial E_r}{\partial r}\right)+
\frac{g_{11}}{\sqrt{-g}}\left(
\frac{\partial E_{\varphi}}{\partial \varphi}\right)\right]\ ,\\
&{}& \left(-\frac{g_{11}}{\sqrt{-g}}\ E_{\varphi}\right)\left[r
\left(\frac{\partial E_r}{\partial t}\right)-
\frac{g_{00}}{\sqrt{-g}}\left(\frac{\partial B}{\partial \varphi}\right)
\right]\ ,\\
&{}& -\frac{g_{11}}{\sqrt{-g}}\left(\frac{\partial
E_{\varphi}}{\partial t}\right)-\left(
\frac{g_{00}}{r^2}-\frac{2}{\ell^2}\right)B-
\frac{g_{00}}{\sqrt{-g}}\left(\frac{\partial B}{\partial t}\right)\ .
\end{eqnarray}
\antiletra
These equation give the three terms in
the expression for $I_{ECS}$.

Now, following \cite{tres} the inverse  temperature as the Euclidean
time period is
\begin{equation}
\beta =\frac{2\pi}{\alpha}
\end{equation}
with $\alpha$ a parameter given by
\begin{equation}
\alpha =\frac{1}{2}\ \frac{df(r)}{dr}\Big|_{r=r_+}\quad , \quad
\alpha \neq 0\ .
\end{equation}
Here the function $f(r)$ is equal to $g_{00}$, and is given by 
\begin{equation}
f(r)=-8MG+\frac{r^2}{\ell^2}
\end{equation}
where $r=r_+$, the event horizon given by 
\begin{equation}
r=r_+=\sqrt{8MG}\ \ell
\end{equation}
where $M,\ G,\ \ell$ are the mass of the black hole, the 
gravitational constant and the cosmological  constant respectively.
Then the temperature, $\beta$, is given as
\begin{equation}
\beta = \frac{\pi \ell}{\sqrt{8MG}}\ \ .
\end{equation}
In general the temperature $T=1/\beta$ defined  in (9) coincides
exactly with the Hawking's temperature for evaporation of black
holes. In our case, if no consideration to topology in the Euclidean
sector is given and if we put off any relation between temperature
and the complex structure of the torus \cite{tres,quatro,cinco} 
the temperature is
\begin{equation}
T_H\sim \frac{\sqrt{M}}{2\pi \ell}\ .
\end{equation}

The total partition function associated with
Einstein-Hilbert-Chern-Simons action is 
\begin{equation}
Z_T\simeq Z_3\cdot Z^{\mbox{{\tiny 
Chern}}}_{\mbox{{\tiny Simons}}}
\end{equation}
where $Z_3$ is the three-dimensional partition function in the saddle
point approximation related to the solution (5) given
in \cite{tres} by 
\begin{equation}
Z_3\simeq e^{\pi^2\ell^2 /2G\beta}
\end{equation}
and $Z_{ECS}$ is the dimensional partition function associated with
the higher derivative Chern-Simons in a gravitational background
\cite{dois} give as 
\begin{equation}
Z_{ECS}\simeq e^{\frac{g_{00}}{\sqrt{-g}}\ BE_r-\left(
\frac{g_{00}}{r^2}-\frac{2}{\ell^2}\right)B}
\end{equation}
For simplicity only two terms from (10) are used. The total partition
function is given as. 
\begin{equation}
Z_T\sim e^{\pi^2\ell^2/2G\beta}
e^{\left(\frac{\pi^2\ell^2}{\beta^2r}-\frac{r}{\ell^2}\right)BE_r}
e^{\left(\frac{\pi^2\ell^2}{\beta^2r}-\frac{3}{\ell^2}\right)B}\ .
\end{equation}

Now the thermodynamical formula for the average energy and the
average entropy $S$ is
\begin{eqnarray}
&{}&M= -\frac{\partial}{\partial \beta}\ (\ln Z_T)\nonumber \\
&{}&S= \ln Z_T-\beta \partial_{\beta}\ln Z_T\ .
\end{eqnarray}

The contribution to entropy is calculated from each  term using (10).
For instance, for
the second term in (10.a) we may write the partition function $Z_T$ as
\begin{equation}
Z_T\sim e^{\pi^2\ell^2/2G\beta}
e^{-\frac{\pi^2\ell^2}{\beta^2}\left(\frac{\partial E_r}{\partial r}\right)
B+\frac{r^2}{\ell^2}\left(\frac{\partial E_r}{\partial r}\right)B}
\end{equation}
Then the average entropy is 
\begin{equation}
S\sim \frac{\pi^2\ell^2}{G\beta}-
\frac{3\pi^2\ell^2}{\beta^2}\left(\frac{\partial E_r}{\partial r}\right)B+
\frac{r^2}{\ell^2}\left(\frac{\partial E_r}{\partial r}\right)B\ .
\end{equation}
One approaching the event horizon $r\rightarrow r_+$ the entropy is 
\begin{equation}
S\sim \frac{\pi^2\ell^2}{G\beta}-
\frac{2\pi^2\ell^2}{\beta^2}\left(\frac{\partial E_r}{\partial r}
\right)_{r=r_+}\cdot B(r=r_+)
\end{equation}
where the first part comes from Einstein-Hilbert action, together
with eq. (6) and the second part comes from extension of Chern-Simons
action in a gravitational background.

Again for simplicity the contribution to entropy only for static
configuration, is considered in eq. (10). The other terms have a non
zero contribution for entropy, in particular a contribution as given
by eq. (20) and eq. (24). The reason why we have a non zero
contribution for entropy 
in the present case is because in constrast with the abelian
Chern-Simons theory for electromagnetic theory or Chern-Simons for
gravitational theory where the energy momentum tensor is zero, here
we find the energy-momentum tensor is not zero and is given as  
\begin{equation}
T^{\mu \nu}_{ECS}=\frac{2}{\sqrt{-g}}\ 
\frac{\delta I_{ECS}}{\delta g_{\mu \nu}}\ .
\end{equation}

The result is written as \cite{dois}
\begin{equation}
T^{\mu \nu}_{ECS}=-m^{-1}\left[\left(\varepsilon^{\mu \alpha
\beta}f^{\nu}+\varepsilon^{\nu \alpha \beta}f^{\mu}
\right)\partial_{\alpha}f_{\beta}-g^{\mu \nu}
\varepsilon^{\alpha \beta \gamma}f_{\alpha}\partial_{\beta}
f_{\gamma}\right]\ ,
\end{equation}
It's interesting to note that if we take the limit such that 
$g_{\mu \nu}\rightarrow \eta_{\mu \nu}$; the equation (4) becomes 
\begin{equation}
f^{\alpha} =\frac{1}{2}\ \varepsilon^{\alpha \mu \nu}F_{\mu \nu}\ ,
\end{equation}
In accordance with \cite{dois}, this cannot be done here since our
metric (5) is a particular case of the anti de-Sitter space.

\section*{Conclusions and Look out:}

\paragraph*{}
In contrast to the abelian Chern-Simons term for the electromagnetic
theory or the Chern-Simons term associated with the gravitational theory
in $D=2+1$ dimensions there is a contribution to the entropy of black holes due to
higher derivative Chern-Simons extensions in a gravitational
background.

Appropriate vector $f^{\alpha}$ for an extension of a topologically
term such as Chern-Simons \cite{dois,seis} is defined and we have shown that
the source of entropy for black holes in $D=2+1$ dimension is the stress tensor
which is not zero $\left(T^{\mu \nu}_{ECS}\neq 0\right)$. 

The entropy using (21) in combination with (17) is different than 
$S=\displaystyle{\frac{A}{4}}$ where $A=2\pi r_+$, the area of
horizon, since here the ``topological contribution'' is not included.

Now, we are considering the contribution to entropy of black holes in $D=2+1$ but
due to non abelian Chern-Simons term such as
\[
S=\frac{k}{4\pi}\int d^3x\varepsilon^{\mu \nu \rho}
\left(\frac{1}{3}\ f^a_{\mu}\partial_{\nu}f^a_{\rho}+
\left(\frac{1}{3!}\right)f^{abc}f^a_{\mu}f^b_{\nu}f^c_{\rho}\right)
\]
where 
\[
f^a_{\mu}=(-g)^{1/2}g_{\mu \alpha}\varepsilon^{\alpha \lambda \gamma}
A^a_{\gamma}\ .
\]
This goal will be hopefully realised in the next letter.

\section*{Aknowledgments:}

\paragraph*{}
I would like to thank the Department of Physics, University of
Alberta for their hospitality. This work was supported by CNPq
(Governamental Brazilian Agencie for Research).

I would like to thank also Dr. Don N. Page for his kindness and attention
with  me at Univertsity of Alberta.

\end{document}